\documentclass[fleqn,usenatbib]{mnras}

\usepackage{newtxtext,newtxmath}

\usepackage[T1]{fontenc}

\DeclareRobustCommand{\VAN}[3]{#2}
\let\VANthebibliography\thebibliography
\def\thebibliography{\DeclareRobustCommand{\VAN}[3]{##3}\VANthebibliography}

\usepackage{graphicx}	
\usepackage{amsmath}	

\title[Cold fronts in Abell 3558]{Two large-scale sloshing cold fronts in the outskirts of the galaxy cluster Abell 3558}

\author[M. S. Mirakhor et al.]{
M. S. Mirakhor,\thanks{Email: msm0033@uah.edu}
S. A. Walker, M. Sundquist and D. Chandra
\\
Department of Physics and Astronomy, The University of Alabama in Huntsville, 301 Sparkman Drive NW, Huntsville, AL 35899, USA
}

\date{Accepted 2023 August 25. Received 2023 July 27; in original form 2023 May 31}

\pubyear{2015}

\begin{document}
\label{firstpage}
\pagerange{\pageref{firstpage}--\pageref{lastpage}}
\maketitle

\begin{abstract}
Previous studies of the massive nearby galaxy cluster Abell 3558 reported a cold front around the cluster core, which is attributed to the sloshing of the core as it responds to the gravitational disturbance created by a past minor merger. Here, using \textit{XMM-Newton} mosaic, we report the detection of two rare large-scale sloshing cold fronts far outside the cooling radius of Abell 3558. One of the detected cold fronts is located 600 kpc from the cluster core to the south-east, while the other is located 1.2 Mpc from the cluster core to the north-west. The latter cold front is one of the most distant cold fronts ever observed in a galaxy cluster. Our findings are in agreement with previous studies that sloshing can extend well beyond the cooling radius, on scales exceeding half the virial radius, suggesting that sloshing is a cluster-wide phenomenon and may affect the cluster's global properties.


\end{abstract}




\section{Introduction}
Cosmological simulations \citep[e.g.][]{Tittley2005,Ascasibar2006,Roediger2011,Roediger2012} predicted that cold fronts observed in the cores of cool-core galaxy clusters are due to the sloshing of low-entropy gas as a consequence of the gravitational disturbance created by infalling galaxies and groups during off-axis minor mergers. These simulations also predicted that the features associated with the central cool gas should propagate outwards over time in a characteristic spiral pattern into lower-pressure regions of galaxy clusters. At these cold fronts, the X-ray surface brightness and gas density drop sharply, whereas the gas temperature rises sharply, with cool gas on the denser (inner) side and hot gas on the rarefied (outer) side, the opposite of what occurs at shock fronts.


In the course of propagating cold fronts from the cluster core to the outskirts, they experience different layers of the intracluster medium (ICM), each of which is dominated by a different physical process. In the core of clusters, the interplay between gas cooling and active galactic nuclei feedback is a dominant process. In the outskirts, the accretion of gas onto the cluster from infalling substructures dominates. Turbulent gas motions due to ongoing mergers increase in the outskirts of the ICM \citep[e.g.][]{Lau2009,Vazza2009}. Consequently, large-scale cold fronts are different from those typically observed in cluster cores, and studying them, therefore, provides a powerful insight into ICM physics of the outskirts of clusters \citep[for recent reviews, see][]{Walker2022book,ZuHone2022book}. 

Large-scale sloshing cold fronts have only been observed in a very small number of galaxy clusters, owing to low surface brightness outside the cooling radius, which is defined as the radius where the cooling time of the ICM is equal to the age of the Universe at a given redshift, approximately 0.01--0.05 of the virial radius\footnote{In this paper, we follow the common practice of referring to $r_{200}$ (the radius within which the mean density is 200 times the critical density of the Universe) as the virial radius.}. Using \textit{XMM--Newton} \citep{Simionescu2012} and \textit{Chandra} \citep{Walker2018NatAs}, a large-scale cold front has been discovered at a radius of 700 kpc from the core of the Perseus cluster to the east. In the western outskirts of Perseus, another large-scale cold front has been identified, reaching out to 1.2 Mpc \citep{Walker2022}. Cold fronts at radii of around half the virial radius have also been observed in a few other clusters: Abell 2142 \citep{Rossetti2013}, RX J2014.8--2430 \citep{Walker2014}, and Abell 1763 \citep{Douglass2018}.

\begin{figure*}
    \centering
    \includegraphics[width=0.9\textwidth]{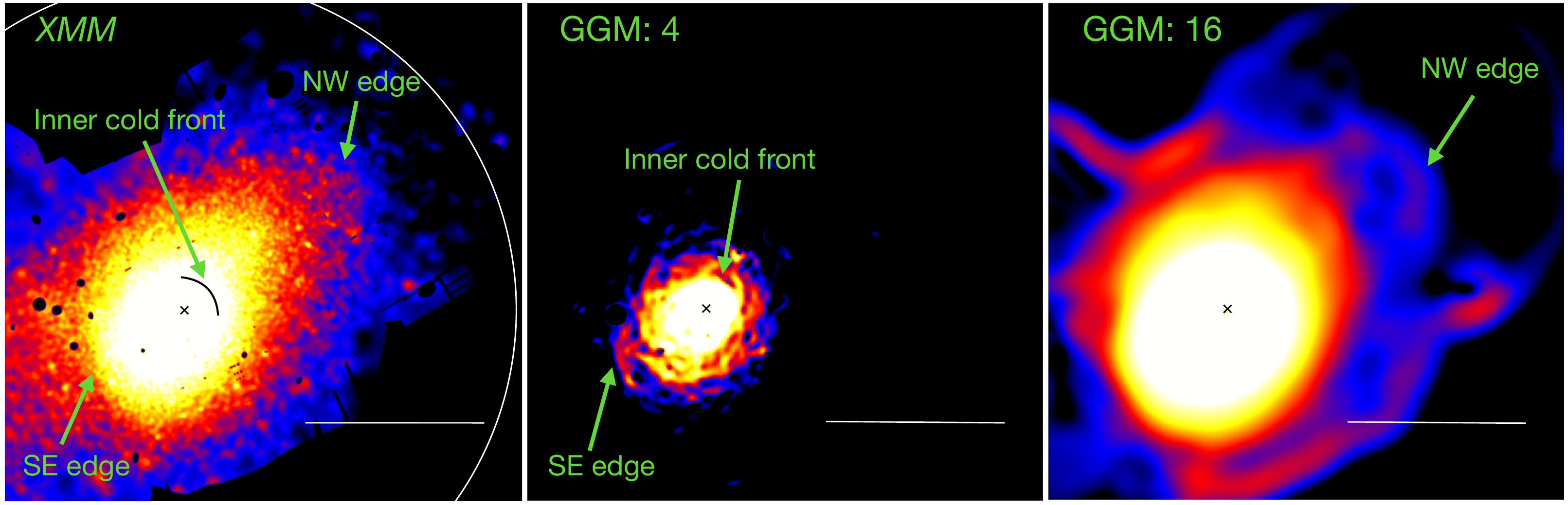}
    \vspace{-2mm}
    \caption{Left-hand panel: Background-subtracted, exposure-corrected \textit{XMM} mosaicked image of the galaxy cluster Abell 3558 in the 0.7--1.2 keV energy band. The white circle marks the location of the $r_{500}$ radius and the black curved line shows the location of the inner cold front. The two newly identified outer edges, one at 600 kpc from the core to the south-east and the other at 1.2 Mpc from the core to the north-west, are highlighted with arrows. Middle and right-hand panels: GGM filtered images with width scales of 4 and 16 pixels, respectively, highlighting the locations of the inner cold front and the outer edges. The cross marks the location of the X-ray peak and the white bar has a length of 15 arcmin (corresponding to 850 kpc). The bright edges to the south and north of the cluster core (right-hand panel) are due to the chip's edge.  }
    \label{fig: A3558_images}
\end{figure*}




The Abell 3558 cluster complex in the central region of the Shapley Supercluster is one of the densest and most massive systems in the local Universe. It consists of three clusters (Abell 3556, Abell 3558 and Abell 3562) and two groups (SC 1327--312 and SC 1329--313). The most massive cluster in the complex is Abell 3558 \citep[$M_{500}=9.8\times10^{14}$ M$_{\odot}$ at the scale radius $r_{500} \approx 1.6$ Mpc;][]{Higuchi2020}, located at redshift $z = 0.048$. \citet{Rossetti2007} identified a cold front at a distance of 100 kpc to the north-west of the core of Abell 3558. The origin of this inner 100 kpc cold front is attributed to large-scale sloshing, induced by a passage of a small group, likely SC 1327--312, or by an off-axis merger with a more massive system \citep{Rossetti2007}. Studies of the core of Abell 3558 with \textit{XMM} and \textit{Chandra} \citep{Rossetti2007,Hudson2010} found that it does not fall into the typical cool core or non-cool core categories. Whilst it has a central surface brightness peak, its central temperature profile slightly drops in the core, leading it to be categorised as a weak cool-core cluster. \citet{Ghizzardi2010} analysed the X-ray data for a sample of galaxy clusters and found that Abell 3558, unlike all clusters that host a sloshing cold front, features a weak entropy gradient, similar to the clusters without cold fronts.



In this paper, based on examining deep \textit{XMM--Newton} data (178.4 ks) of Abell 3558, we report the discovery of two rare large-scale cold fronts, reaching out to very large radii. One of them is located $\sim$ 600 kpc to the south-east of the core, while the other is located $\sim$1.2 Mpc to the north-west. Previous work by \citet{Rossetti2007} used much shallower \textit{XMM} (44.6 ks) and \textit{Chandra} (14.4 ks) observations, only covering the central region of Abell 3558 in which the outer northwestern edge was not covered. Although the outer southeastern edge was covered by this shallow data, this edge remained undetected, probably due to the brevity of the data used by \citet{Rossetti2007}.


Throughout the paper, we adopt a $\Lambda$CDM cosmology with $\Omega_{\rm{\Lambda}}=0.7$, $\Omega_{\rm{m}}=0.3$, and $H_0=70$~km s$^{-1}$ Mpc$^{-1}$. At the redshift of Abell 3558, 1 arcsec corresponds to 0.947 kpc. Using Navarro-Frenk-White (NFW) profile \citep{Navarro1997}, \citet{Higuchi2020} found that the scale radius of Abell 3558 at a density 200 times the critical density of the Universe is $r_{200}=2.3$ Mpc, and this implies that the scale radius $r_{500}=1.6$ Mpc, using the relation $r_{500}=0.7r_{200}$ obtained assuming a NFW profile with a concentration parameter of 4. 


\section{Data reduction}
\label{sec: data}
We use five archival \textit{XMM--Newton} observations for a total unfiltered exposure of 178.4 ks (filtered exposure time is 97.5 ks). One of the observations (Obs. ID: 0107260101) was carried out in 2002 for an unfiltered exposure of 44.6 ks, covering the central region of Abell 3558. The other observations (Obs. ID: 0601980101, 0601980301, 0651590201, 0744930201) were conducted between 2009 and 2014 for an unfiltered exposure of 133.8 ks, covering the southeastern and northwestern parts of the cluster.

The data were reduced using the \textit{XMM--Newton} Science Analysis System, \textit{XMM}-SAS version 19.1 and the Current Calibration Files database, following the procedures illustrated in the Extended Source Analysis Software manual \citep[ESAS;][]{snowden2014cookbook}, as is also done in \citet{Mirakhor2020high}, \citet{Mirakhor2021exploring}, \citet{Mirakhor2021Virgo}, and \citet{Mirakhor2022A2029}. We processed the data by running the \textsc{epchain} and \textsc{emchain} tasks, followed by the tasks \textsc{mos-filter} and \textsc{pn-filter}. The images and exposure maps were extracted in the 0.7--1.2 keV, using \textsc{mos-spectra} and \textsc{pn-spectra}, while particle background images were produced using \textsc{mos-back} and \textsc{mos-back}. This energy range was chosen in order to minimise the high-energy-particle background, which rises significantly at low and high energies. The MOS detectors were examined for CCDs in anomalous data and affected CCDs have been removed from the analysis. The data were also examined for any residual soft proton contamination using the \textsc{proton} task. Point sources in the field were detected and masked by running the ESAS source-detection tool \textsc{cheese}. As is done in \citet{Rossetti2013}, we excluded the data of the PN detector, since its background was found to be less predictable than for the MOS detectors.


In the left panel of Fig. \ref{fig: A3558_images}, we show the background-subtracted, exposure-corrected mosaic image of Abell 3558. The location of the inner cold front \citep{Rossetti2007} is marked in the figure. There are two visible edges, one at 600 kpc from the core to the south-east, and the other at 1.2 Mpc from the core to the north-west. To highlight the surface brightness edges in the X-ray image and to show how they are related to the inner cold front, we applied the Gaussian Gradient Magnitude (GGM) filter \citep{Sanders2016b,Walker2016}. The GGM filter convolves the image with a Gaussian kernel and determines the gradient on the kernel's spatial scale. The middle and right-hand panels of Fig. \ref{fig: A3558_images} show the GGM-filtered images with width scales of 4 and 16 pixels. The GGM-filtered images show significant gradients in the surface brightness across both the southeastern and northwestern edges at a level of 6--7$\sigma$. In addition to these two significant gradients, a few other gradients in the surface brightness can also be seen in the filtered images of Abell 3558, but they are less significant (1--3$\sigma$), and thus not considered for further analysis. In Sections \ref{sec: SE} and \ref{sec: NW}, we probe the spatial and spectral properties of both edges in order to determine whether  they are cold or shock fronts.

\section{Southeastern edge}
\label{sec: SE}
\subsection{Imaging analysis}
\label{sec: imaging_SE}
As shown in Fig. \ref{fig: A3558_images}, there is a sharp edge in the X-ray surface brightness of Abell 3558 at $\sim$600 kpc from its centre to the south-east. To probe the properties of this edge, we extracted the surface brightness profile in a sector along the southeastern direction with position angles 140--170 deg, where the angles are defined counterclockwise with respect to the Right Ascension axis. These angles were chosen as a maximum gradient in surface brightness was found in that direction, following the X-ray isophotes of Abell 3558 closely and using elliptical annuli. A local background far beyond the core of the cluster ($\sim$30 arcmin from the core) to the north-west was subtracted from the surface brightness of the cluster.

In Fig. \ref{fig: SE_edge}, we show the surface brightness profile along the southeastern direction, where the outer 600 kpc edge is identified. To determine a possible density discontinuity across this edge, we fitted the surface brightness profile with a broken power-law model projected along the line of sight \citep{Markevitch2007}, as is also done, for example, in \citet{Botteon2018} and \citet{Mirakhor2023}. The best-fitting model is shown as a solid blue line in Fig. \ref{fig: SE_edge} (upper panel), and yields a density drop by a factor of $1.8_{-0.1}^{+0.1}$ at a distance of 9.7 arcmin (corresponding roughly to 600 kpc) from the surface brightness peak, far beyond the cooling radius of the cluster.

\subsection{Spectral analysis}
\label{sec: Spectral_SE}
In order to confirm whether the outer edge to the south-east of the core is a cold or shock front, we need to carry out the spectral analysis along the southeastern direction. We, therefore, extracted the spectra and response files from elliptical annuli following the X-ray isophotes closely using the ESAS tasks \textsc{mos-spectra} and \textsc{mos-back}. Because of the low surface brightness, particularly at the outer part of the edge, it is essential to model all of the various background components in order to obtain reliable measurements of the relevant parameters.

We followed the procedure described in the ESAS cookbook, as is also done in \citet{Bulbul2012} and \citet{Walker2013_CentaurusXMM}. The background model consists of three main components: the instrumental background, the cosmic X-ray background, and the solar wind charge exchange background. The instrumental background consists of instrumental fluorescent lines and soft proton background. The Al K$\alpha$ and Si K$\alpha$  instrumental fluorescent lines at energies $E = 1.49$ keV and $E = 1.75$ keV in the MOS detectors were modelled with two Gaussian components. Residual soft proton contamination, which may be present, was modelled with a broken power-law model, as in \citet{Leccardi2008}, using diagonal spectral redistribution matrices provided by the ESAS Current Calibration Files database. The power-law indexes were left free to vary between 0.2 and 1.4. 

The X-ray cosmic background is well known but has to be modelled explicitly since it is a significant background in all directions and on a broad range of energies. It consists of these spectral components: unabsorbed thermal component at $E\sim$0.1 keV representing the emission from the Local Hot Bubble; absorbed thermal component at $E\sim$ 0.1 keV accounting for the cooler halo emission; absorbed thermal component at $E\sim$0.25 keV representing emission from the intergalactic medium or the hotter halo; and an absorbed power law with photon index fixed at 1.4 representing emission from the unresolved background of cosmological sources. In order to set a tight constraint on the cosmic background, we used the \textit{ROSAT} All-Sky Survey (RASS) spectrum from the \textsc{heasarc} X-ray background tool. This spectrum was extracted in an annulus surrounding the Abell 3558 cluster beyond its virial radius, assuming that it represents reasonably well the cosmic background in the direction of the cluster. We allowed for cross-calibration uncertainty between \textit{ROSAT} and \textit{XMM}, and found that the cross-calibration difference between the two instruments is very small (< 3 per cent). For the potential solar wind charge exchange emission, two Gaussian components were used to model the O \textsc{vii} line at 0.56 keV, and the O \textsc{viii} line at 0.65 keV, with widths set to zero.

\begin{figure}
    \centering
    \includegraphics[width=0.9\columnwidth]{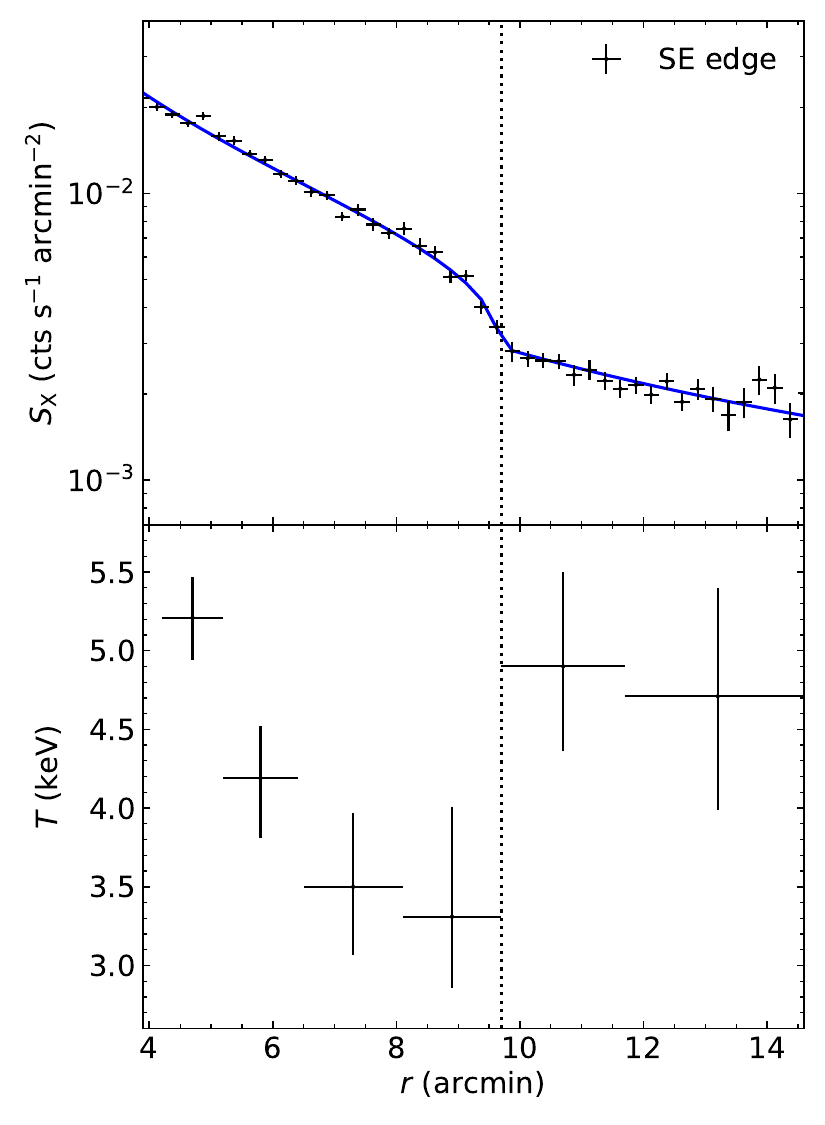}
    \vspace{-4mm}
    \caption{Upper panel: Surface brightness profile along the southeastern direction, where the outer 600 kpc edge is identified. The best fit to the data using a broken power-law model is shown as a solid blue line. Lower panel: Temperature profile along the southeastern direction as noted above. The dotted vertical line show the location of the edge.  } 
    \label{fig: SE_edge}
\end{figure}

The \textit{XMM} source spectrum was fitted with a single-temperature model, consisting of an APEC thermal component \citep{Smith2001}, with the redshift fixed at $z=0.048$ and the Galactic absorption fixed at $N_{\rm{H}}=3.66 \times 10^{20}$ cm$^{-2}$. The temperature and abundance were left free to vary. Using the standard X-ray spectral fitting package \citep[\textsc{xspec;}][]{Arnaud1996} and the abundance table of \citet{anders1989abundances}, the spectral fitting was carried out in the 0.4--11 keV band. The abundance in an approximate range of 0.2--0.3 $Z_{\odot}$ was obtained. In the lower panel of Fig.  \ref{fig: SE_edge}, we show the temperature profile along the southeastern direction, where the outer 600 kpc edge is identified. There is a jump in the gas temperature at a level of 1.8$\sigma$ from $3.3_{-0.5}^{+0.7}$ keV inside the edge to $4.9_{-0.5}^{+0.6}$ keV outside of it. Given that the gas temperature is lower inside the edge than its outside, we, therefore, conclude that this edge is a cold front.

\section{Northwestern edge}
\label{sec: NW}
\begin{figure}
    \centering
    \includegraphics[width=0.9\columnwidth]{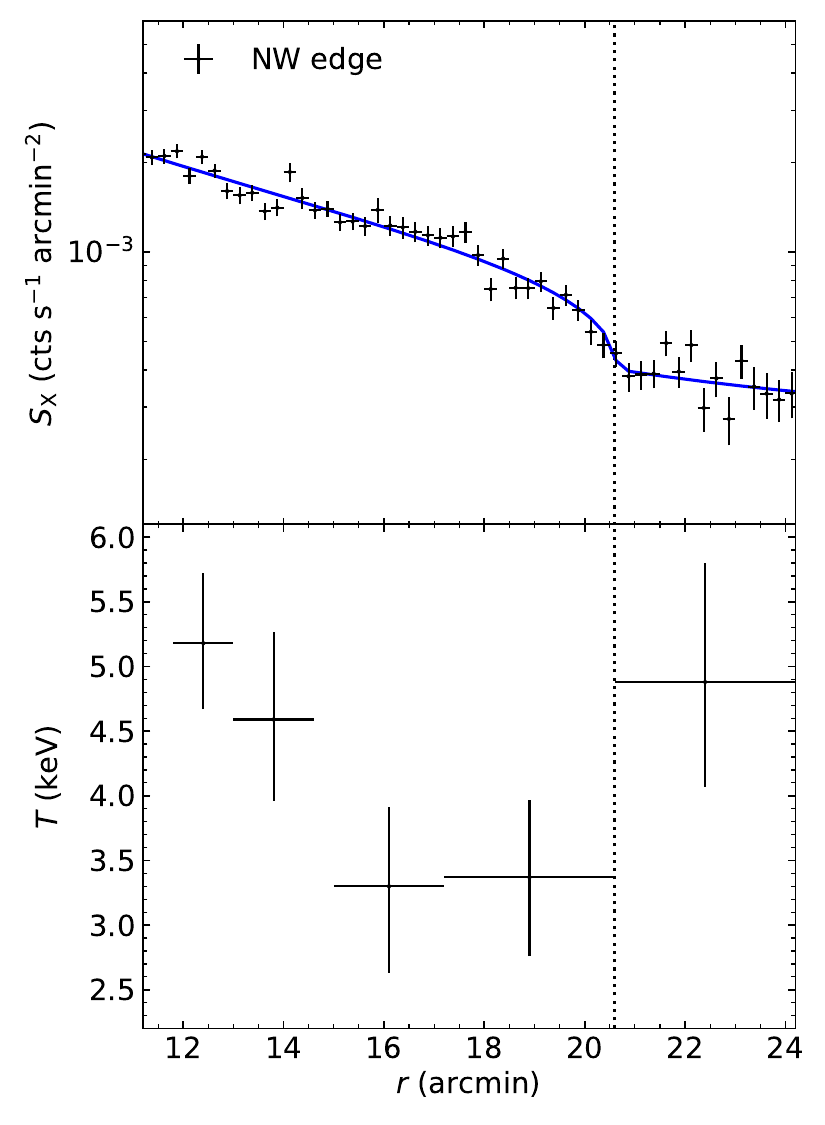}
    \vspace{-4mm}
    \caption{Upper panel: Surface brightness profile along the northwestern direction, where the outer north-west edge is identified. The best fit to the data using a broken power-law model is shown as a solid blue line. Lower panel: Temperature profile along the southeastern direction as noted above. The dotted vertical line show the location of the edge.  } 
    \label{fig: NW_edge}
\end{figure}

As is done in Section \ref{sec: imaging_SE}, we extracted the surface brightness profile in elliptical annuli along the northwestern direction with position angles 165-190 deg. Similar to the southeastern edge, these angles were chosen as we found a maximum gradient in surface brightness along this angular range, following closely the X-ray isophotes and using elliptical annuli. The upper panel of Fig. \ref{fig: NW_edge} shows the surface brightness profile along the northwestern direction, where the outer edge at 1.2 Mpc is identified. The best fit to the surface brightness profile using a broken power-law model (blue solid line) yields a sharp drop in the gas density by a factor of $2.2_{-0.2}^{+0.2}$ at a distance of 20.7 arcmin (corresponding roughly to 1.2 Mpc) from the surface brightness peak. This edge exceeds half of the virial radius of Abell 3558 \citep[$r_{200} = 2.3$ Mpc;][]{Higuchi2020}.

To determine the nature of the outer northwestern discontinuity, we carried out a spectral analysis along this direction, following the same procedure described in Section \ref{sec: Spectral_SE}. Our best-fitting temperature profile is shown in the lower panel of Fig. \ref{fig: NW_edge}. At this outer edge, we note that the gas temperature jumps at a level of 1.1$\sigma$ from $3.4_{-0.6}^{+0.6}$ keV inside the edge to $4.9_{-0.8}^{+0.9}$ keV outside of it, indicating that the edge is a cold front. Since it locates 1.2 Mpc from the cluster core, the outermost northwestern edge is one of the most distant cold fronts ever observed in a galaxy cluster. 

It is worth mentioning that the associated errors with the temperature measurements are statistical errors, but there might be some sources of systematic errors such as systematics in the background estimation particularly at the outer part of the edge, which might change our conclusions regarding the nature of the edge. To estimate the possible role of systematic errors in our temperature measurements, we allowed the background parameters (normalisations) to vary within $\pm 5$ per cent of the best-fitting values, as is done in \citet{Ghirardini2018}. We simulated 1000 different realisations of the background, assuming that the fluctuations follow a Gaussian distribution. By subtracting each realisations of the generated backgrounds from the source spectrum and fitting with an APEC thermal model, we found a systematic error of about 6 per cent, suggesting that the dominant error is a statistical error. This, in turn, suggests that the northwestern edge is likely a cold front, although the temperature difference is only at a level of 1$\sigma$. Alternatively, if we assume that this edge is a shock front with a density jump of 2.2 and an inner temperature of 3.4 keV as estimated above, we predict an outer temperature of 1.7 keV using the Rankine-Hugoniot condition. By adopting this temperature for the outside of the edge, we find a difference in C-statistic of about 31.4 relative to the fit with a temperature of 4.9 keV, which implies a rejection of this hypothesis at a level of 5.8$\sigma$. We, therefore, conclude this edge is a cold front.

\section{Discussion and Summary}
\label{sec: discussion}
Our detailed analysis of the \textit{XMM} observations of the Abell 3558 cluster has revealed two abrupt surface brightness discontinuities at large-scale radii outside the cluster core that have not been reported previously. One of the discontinuities is located 600 kpc from the cluster core to the south-east and the other is located 1.2 Mpc (exceeds half the virial radius) from the cluster core to the north-west. The spectral analysis of the regions inside and outside the identified surface brightness discontinuities indicates that these features are large-scale cold fronts. These two newly detected cold fronts, together with the inner cold front reported in \citet{Rossetti2007}, make Abell 3558 one of only a few clusters that host three or more cold fronts. All these galaxy clusters are massive hot systems that show signs of merger activity. It is possible that these systems underwent a merger event in the distant past with a certain impact parameter which resulted in these distant and rare features. There is also a possibility that these features are rare due to the fact of low surface brightness outside the core and limitation of the sensitivity of current X-ray instruments that do not allow to detect these features more often, or the observation is not long enough in other systems to detect them.

In agreement with the other clusters in which large-scale cold fronts have been observed, Abell 3558 features an outwardly spiralling pattern of concentric cold fronts on opposite sides of the cluster, connecting with the cold front in the core. This pattern of cold fronts on opposite sides of the cluster, at increasing radius, may indicate that the outer cold fronts in Abell 3558 formed from the same merger that has induced sloshing in the cluster core. Simulations \citep{Walker2018NatAs} predicted that the time evolution of the cold front radius is well fit with a linear relation, with the cold front rising at constant speed. By extrapolating this linear best fit to the radius of our observed outer cold front to the north-west, we conservatively estimate an age of $\sim 8.5$ Gyr for this cold front from the moment of closest approach between a possible subsystem and Abell 3558. Simulations tailored specifically to the Abell 3558 cluster will be the subject of future work.

Cosmological simulations \citep[e.g.][]{Zuhone2016review,Brzycki2019} indicate that the formation of large-scale cold fronts requires low-mass ratio mergers, which can move cold gas outward from the cluster core to larger radii. These simulations also showed that the large impact parameter is also essential to produce large cold fronts since a lower impact parameter merger would destroy the core. This is the case for Abell 3558 where its driven thermodynamic properties \citep{Rossetti2007} indicate that the cluster has been perturbed in the past by a low-mass merger, likely SC 1327--312, without completely destroying its core.



\begin{figure}
    \centering
    \includegraphics[width=0.9\columnwidth]{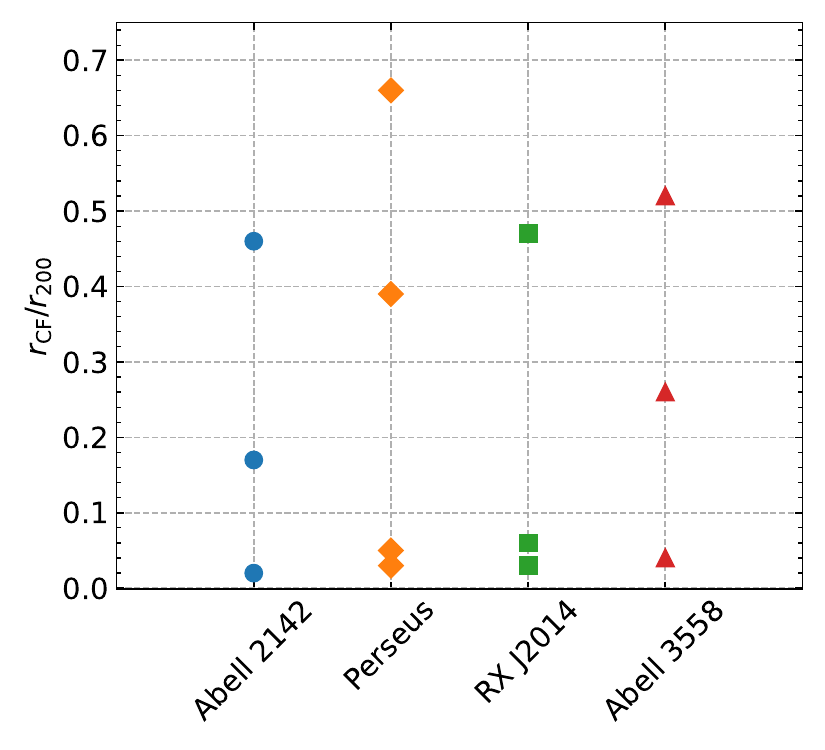}
    \vspace{-6mm}
    \caption{Comparing the locations of the cold fronts in Abell 3558 with those in Abell 2142, the Perseus cluster, and RX J2014.8--2430. We note that the $r_{200}$ values used for the rescaling are not estimated from the same analysis and technique, thus the locations of the cold fronts might be slightly biased.}
    \label{fig: comparison}
\end{figure}

Recently, using radio data from ASKAP and MeerKAT, \citet{Venturi2022} detected diffuse radio emission at the centre of Abell 3558. The detected radio source is small in size ($200\times400$ kpc), and it has a very steep spectrum ($\alpha = -2.3$) with an extremely low radio power ($P_{{1283}\,\rm{MHz}} = 6.85 \times 10^{22}$ W Hz$^{-1}$). Interestingly, the location of this radio emission coincides with the region of low-entropy gas within the envelope of the inner 100 kpc cold front to the north-west and the outer 600 kpc cold front to the south-east. Although it is unclear whether this diffuse emission should be classified as a radio halo or a radio mini-halo \citep{Venturi2022}, its properties support the prediction of cosmological simulations \citep[e.g.][]{ZuHone2013} that its origin is the same gas-sloshing mechanism responsible for the formation of the sloshing cold fronts in the cluster. The case of Abell 3558 is similar to Abell 2142, another galaxy cluster that hosts a large-scale cold front, where large-scale sloshing was invoked as a possible re-acceleration mechanism for the origin of the Mpc-scale radio emission in the cluster \citep{Rossetti2013}. 

The location of the outermost northwestern cold front is 1.2 Mpc from the cluster core, making it one of the most distant cold fronts ever observed in a cluster. It is located at a greater distance from the cluster core than the eastern cold front in the Perseus cluster \citep[0.7 Mpc;][]{Simionescu2012}, the outermost cold front in Abell 2142 \citep[1.0 Mpc;][]{Rossetti2013}, and the southern cold front in RX J2014.8--2430 \citep[0.8 Mpc;][]{Walker2014}. It is located at the same distance from the cluster core as that observed in the western outskirts of Perseus \citep[1.2 Mpc;][]{Walker2022}. 

In Fig. \ref{fig: comparison}, we compare the locations of the cold fronts in Abell 3558 with those observed in Perseus \citep{Simionescu2012,Walker2022}, Abell 2142 \citep{Rossetti2013}, and RX J2014.8--2430 \citep{Walker2014}. We compare them as a fraction of the scale radius $r_{200}$ to account for the different masses of the clusters. The location of the outermost northwestern cold front in Abell 3558 corresponds to $0.52 r_{200}$. In terms of the fraction of the $r_{200}$ radius, it is located at a greater distance from the core than most of the large-scale sloshing cold fronts observed in clusters, with the exception of the 1.2 Mpc cold front located in the western outskirts of Perseus.



\section*{ACKNOWLEDGEMENTS}
We thank the referee for their helpful comments that improved the paper. This work is based on observations obtained with \textit{XMM--Newton}, an ESA science mission with instruments and contributions directly funded by ESA Member States and NASA.

\section*{Data Availability}
The \textit{XMM--Newton} Science Archive (XSA) stores the archival data used in this paper, from which the data are publicly available for download.



\bibliographystyle{mnras}
\bibliography{A3558} 

\begin{thebibliography}{}
\makeatletter
\relax
\def\mn@urlcharsother{\let\do\@makeother \do\$\do\&\do\#\do\^\do\_\do\%\do\~}
\def\mn@doi{\begingroup\mn@urlcharsother \@ifnextchar [ {\mn@doi@}
  {\mn@doi@[]}}
\def\mn@doi@[#1]#2{\def\@tempa{#1}\ifx\@tempa\@empty \href
  {http://dx.doi.org/#2} {doi:#2}\else \href {http://dx.doi.org/#2} {#1}\fi
  \endgroup}
\def\mn@eprint#1#2{\mn@eprint@#1:#2::\@nil}
\def\mn@eprint@arXiv#1{\href {http://arxiv.org/abs/#1} {{\tt arXiv:#1}}}
\def\mn@eprint@dblp#1{\href {http://dblp.uni-trier.de/rec/bibtex/#1.xml}
  {dblp:#1}}
\def\mn@eprint@#1:#2:#3:#4\@nil{\def\@tempa {#1}\def\@tempb {#2}\def\@tempc
  {#3}\ifx \@tempc \@empty \let \@tempc \@tempb \let \@tempb \@tempa \fi \ifx
  \@tempb \@empty \def\@tempb {arXiv}\fi \@ifundefined
  {mn@eprint@\@tempb}{\@tempb:\@tempc}{\expandafter \expandafter \csname
  mn@eprint@\@tempb\endcsname \expandafter{\@tempc}}}

\bibitem[\protect\citeauthoryear{{Anders} \& {Grevesse}}{{Anders} \&
  {Grevesse}}{1989}]{anders1989abundances}
{Anders} E.,  {Grevesse} N.,  1989, \mn@doi [\gca]
  {10.1016/0016-7037(89)90286-X}, \href
  {https://ui.adsabs.harvard.edu/abs/1989GeCoA..53..197A} {53, 197}

\bibitem[\protect\citeauthoryear{{Arnaud}}{{Arnaud}}{1996}]{Arnaud1996}
{Arnaud} K.~A.,  1996, in {Jacoby} G.~H.,  {Barnes} J.,  eds,  Astronomical
  Society of the Pacific Conference Series Vol. 101, Astronomical Data Analysis
  Software and Systems V. p.~17

\bibitem[\protect\citeauthoryear{{Ascasibar} \& {Markevitch}}{{Ascasibar} \&
  {Markevitch}}{2006}]{Ascasibar2006}
{Ascasibar} Y.,  {Markevitch} M.,  2006, \mn@doi [\apj] {10.1086/506508}, \href
  {http://adsabs.harvard.edu/abs/2006ApJ...650..102A} {650, 102}

\bibitem[\protect\citeauthoryear{{Botteon}, {Gastaldello}  \&
  {Brunetti}}{{Botteon} et~al.}{2018}]{Botteon2018}
{Botteon} A.,  {Gastaldello} F.,   {Brunetti} G.,  2018, \mn@doi [\mnras]
  {10.1093/mnras/sty598}, \href
  {https://ui.adsabs.harvard.edu/abs/2018MNRAS.476.5591B} {476, 5591}

\bibitem[\protect\citeauthoryear{{Brzycki} \& {ZuHone}}{{Brzycki} \&
  {ZuHone}}{2019}]{Brzycki2019}
{Brzycki} B.,  {ZuHone} J.,  2019, \mn@doi [\apj] {10.3847/1538-4357/ab3983},
  \href {https://ui.adsabs.harvard.edu/abs/2019ApJ...883..118B} {883, 118}

\bibitem[\protect\citeauthoryear{{Bulbul}, {Smith}, {Foster}, {Cottam},
  {Loewenstein}, {Mushotzky}  \& {Shafer}}{{Bulbul} et~al.}{2012}]{Bulbul2012}
{Bulbul} G.~E.,  {Smith} R.~K.,  {Foster} A.,  {Cottam} J.,  {Loewenstein} M.,
  {Mushotzky} R.,   {Shafer} R.,  2012, \mn@doi [\apj]
  {10.1088/0004-637X/747/1/32}, \href
  {https://ui.adsabs.harvard.edu/abs/2012ApJ...747...32B} {747, 32}

\bibitem[\protect\citeauthoryear{{Douglass}, {Blanton}, {Randall}, {Clarke},
  {Edwards}, {Sabry}  \& {ZuHone}}{{Douglass} et~al.}{2018}]{Douglass2018}
{Douglass} E.~M.,  {Blanton} E.~L.,  {Randall} S.~W.,  {Clarke} T.~E.,
  {Edwards} L.~O.~V.,  {Sabry} Z.,   {ZuHone} J.~A.,  2018, \mn@doi [\apj]
  {10.3847/1538-4357/aae9e7}, \href
  {https://ui.adsabs.harvard.edu/abs/2018ApJ...868..121D} {868, 121}

\bibitem[\protect\citeauthoryear{{Ghirardini}, {Ettori}, {Eckert}, {Molendi},
  {Gastaldello}, {Pointecouteau}, {Hurier}  \& {Bourdin}}{{Ghirardini}
  et~al.}{2018}]{Ghirardini2018}
{Ghirardini} V.,  {Ettori} S.,  {Eckert} D.,  {Molendi} S.,  {Gastaldello} F.,
  {Pointecouteau} E.,  {Hurier} G.,   {Bourdin} H.,  2018, \mn@doi [\aap]
  {10.1051/0004-6361/201731748}, \href
  {https://ui.adsabs.harvard.edu/abs/2018A&A...614A...7G} {614, A7}

\bibitem[\protect\citeauthoryear{{Ghizzardi}, {Rossetti}  \&
  {Molendi}}{{Ghizzardi} et~al.}{2010}]{Ghizzardi2010}
{Ghizzardi} S.,  {Rossetti} M.,   {Molendi} S.,  2010, \mn@doi [\aap]
  {10.1051/0004-6361/200912496}, \href
  {https://ui.adsabs.harvard.edu/abs/2010A&A...516A..32G} {516, A32}

\bibitem[\protect\citeauthoryear{{Higuchi}, {Okabe}, {Merluzzi}, {Haines},
  {Busarello}, {Grado}  \& {Mercurio}}{{Higuchi} et~al.}{2020}]{Higuchi2020}
{Higuchi} Y.,  {Okabe} N.,  {Merluzzi} P.,  {Haines} C.~P.,  {Busarello} G.,
  {Grado} A.,   {Mercurio} A.,  2020, \mn@doi [\mnras]
  {10.1093/mnras/staa1766}, \href
  {https://ui.adsabs.harvard.edu/abs/2020MNRAS.497...52H} {497, 52}

\bibitem[\protect\citeauthoryear{{Hudson}, {Mittal}, {Reiprich}, {Nulsen},
  {Andernach}  \& {Sarazin}}{{Hudson} et~al.}{2010}]{Hudson2010}
{Hudson} D.~S.,  {Mittal} R.,  {Reiprich} T.~H.,  {Nulsen} P.~E.~J.,
  {Andernach} H.,   {Sarazin} C.~L.,  2010, \mn@doi [\aap]
  {10.1051/0004-6361/200912377}, \href
  {https://ui.adsabs.harvard.edu/abs/2010A&A...513A..37H} {513, A37}

\bibitem[\protect\citeauthoryear{{Lau}, {Kravtsov}  \& {Nagai}}{{Lau}
  et~al.}{2009}]{Lau2009}
{Lau} E.~T.,  {Kravtsov} A.~V.,   {Nagai} D.,  2009, \mn@doi [\apj]
  {10.1088/0004-637X/705/2/1129}, \href
  {http://adsabs.harvard.edu/abs/2009ApJ...705.1129L} {705, 1129}

\bibitem[\protect\citeauthoryear{{Leccardi} \& {Molendi}}{{Leccardi} \&
  {Molendi}}{2008}]{Leccardi2008}
{Leccardi} A.,  {Molendi} S.,  2008, \mn@doi [\aap]
  {10.1051/0004-6361:200809538}, \href
  {http://adsabs.harvard.edu/abs/2008A%26A...486..359L} {486, 359}

\bibitem[\protect\citeauthoryear{{Markevitch} \& {Vikhlinin}}{{Markevitch} \&
  {Vikhlinin}}{2007}]{Markevitch2007}
{Markevitch} M.,  {Vikhlinin} A.,  2007, \mn@doi [\physrep]
  {10.1016/j.physrep.2007.01.001}, \href
  {http://adsabs.harvard.edu/abs/2007PhR...443....1M} {443, 1}

\bibitem[\protect\citeauthoryear{{Mirakhor} \& {Walker}}{{Mirakhor} \&
  {Walker}}{2020}]{Mirakhor2020high}
{Mirakhor} M.~S.,  {Walker} S.~A.,  2020, \mn@doi [\mnras]
  {10.1093/mnras/staa2204}, \href
  {https://ui.adsabs.harvard.edu/abs/2020MNRAS.497.3943M} {497, 3943}

\bibitem[\protect\citeauthoryear{{Mirakhor} \& {Walker}}{{Mirakhor} \&
  {Walker}}{2021}]{Mirakhor2021Virgo}
{Mirakhor} M.~S.,  {Walker} S.~A.,  2021, \mn@doi [\mnras]
  {10.1093/mnras/stab1768}, \href
  {https://ui.adsabs.harvard.edu/abs/2021MNRAS.506..139M} {506, 139}

\bibitem[\protect\citeauthoryear{{Mirakhor} et~al.,}{{Mirakhor}
  et~al.}{2021}]{Mirakhor2021exploring}
{Mirakhor} M.~S.,  et~al., 2021, \mn@doi [\mnras] {10.1093/mnras/staa3404},
  \href {https://ui.adsabs.harvard.edu/abs/2021MNRAS.500.2503M} {500, 2503}

\bibitem[\protect\citeauthoryear{{Mirakhor}, {Walker}  \& {Runge}}{{Mirakhor}
  et~al.}{2022}]{Mirakhor2022A2029}
{Mirakhor} M.~S.,  {Walker} S.~A.,   {Runge} J.,  2022, \mn@doi [\mnras]
  {10.1093/mnras/stab2979}, \href
  {https://ui.adsabs.harvard.edu/abs/2022MNRAS.509.1109M} {509, 1109}

\bibitem[\protect\citeauthoryear{{Mirakhor}, {Walker}  \& {Runge}}{{Mirakhor}
  et~al.}{2023}]{Mirakhor2023}
{Mirakhor} M.~S.,  {Walker} S.~A.,   {Runge} J.,  2023, \mn@doi [\mnras]
  {10.1093/mnras/stad1088}, \href
  {https://ui.adsabs.harvard.edu/abs/2023MNRAS.522.2105M} {522, 2105}

\bibitem[\protect\citeauthoryear{{Navarro}, {Frenk}  \& {White}}{{Navarro}
  et~al.}{1997}]{Navarro1997}
{Navarro} J.~F.,  {Frenk} C.~S.,   {White} S. D.~M.,  1997, \mn@doi [\apj]
  {10.1086/304888}, \href
  {https://ui.adsabs.harvard.edu/abs/1997ApJ...490..493N} {490, 493}

\bibitem[\protect\citeauthoryear{{Roediger}, {Br{\"u}ggen}, {Simionescu},
  {B{\"o}hringer}, {Churazov}  \& {Forman}}{{Roediger}
  et~al.}{2011}]{Roediger2011}
{Roediger} E.,  {Br{\"u}ggen} M.,  {Simionescu} A.,  {B{\"o}hringer} H.,
  {Churazov} E.,   {Forman} W.~R.,  2011, \mn@doi [\mnras]
  {10.1111/j.1365-2966.2011.18279.x}, \href
  {http://adsabs.harvard.edu/abs/2011MNRAS.413.2057R} {413, 2057}

\bibitem[\protect\citeauthoryear{{Roediger}, {Lovisari}, {Dupke}, {Ghizzardi},
  {Br{\"u}ggen}, {Kraft}  \& {Machacek}}{{Roediger}
  et~al.}{2012}]{Roediger2012}
{Roediger} E.,  {Lovisari} L.,  {Dupke} R.,  {Ghizzardi} S.,  {Br{\"u}ggen} M.,
   {Kraft} R.~P.,   {Machacek} M.~E.,  2012, \mn@doi [\mnras]
  {10.1111/j.1365-2966.2011.20287.x}, \href
  {https://ui.adsabs.harvard.edu/abs/2012MNRAS.420.3632R} {420, 3632}

\bibitem[\protect\citeauthoryear{{Rossetti}, {Ghizzardi}, {Molendi}  \&
  {Finoguenov}}{{Rossetti} et~al.}{2007}]{Rossetti2007}
{Rossetti} M.,  {Ghizzardi} S.,  {Molendi} S.,   {Finoguenov} A.,  2007,
  \mn@doi [\aap] {10.1051/0004-6361:20054621}, \href
  {https://ui.adsabs.harvard.edu/abs/2007A&A...463..839R} {463, 839}

\bibitem[\protect\citeauthoryear{{Rossetti}, {Eckert}, {De Grandi},
  {Gastaldello}, {Ghizzardi}, {Roediger}  \& {Molendi}}{{Rossetti}
  et~al.}{2013}]{Rossetti2013}
{Rossetti} M.,  {Eckert} D.,  {De Grandi} S.,  {Gastaldello} F.,  {Ghizzardi}
  S.,  {Roediger} E.,   {Molendi} S.,  2013, \mn@doi [\aap]
  {10.1051/0004-6361/201321319}, \href
  {http://adsabs.harvard.edu/abs/2013A%26A...556A..44R} {556, A44}

\bibitem[\protect\citeauthoryear{{Sanders}, {Fabian}, {Russell}, {Walker}  \&
  {Blundell}}{{Sanders} et~al.}{2016}]{Sanders2016b}
{Sanders} J.~S.,  {Fabian} A.~C.,  {Russell} H.~R.,  {Walker} S.~A.,
  {Blundell} K.~M.,  2016, \mn@doi [\mnras] {10.1093/mnras/stw1119}, \href
  {http://adsabs.harvard.edu/abs/2016MNRAS.460.1898S} {460, 1898}

\bibitem[\protect\citeauthoryear{{Simionescu} et~al.,}{{Simionescu}
  et~al.}{2012}]{Simionescu2012}
{Simionescu} A.,  et~al., 2012, \mn@doi [\apj] {10.1088/0004-637X/757/2/182},
  \href {http://adsabs.harvard.edu/abs/2012ApJ...757..182S} {757, 182}

\bibitem[\protect\citeauthoryear{{Smith}, {Brickhouse}, {Liedahl}  \&
  {Raymond}}{{Smith} et~al.}{2001}]{Smith2001}
{Smith} R.~K.,  {Brickhouse} N.~S.,  {Liedahl} D.~A.,   {Raymond} J.~C.,  2001,
  \mn@doi [\apjl] {10.1086/322992}, \href
  {http://adsabs.harvard.edu/abs/2001ApJ...556L..91S} {556, L91}

\bibitem[\protect\citeauthoryear{Snowden \& Kuntz}{Snowden \&
  Kuntz}{2014}]{snowden2014cookbook}
Snowden S.~L.,  Kuntz K.~D.,  2014, Cookbook for analysis procedures for
  XMM-Newton EPIC observations of extended objects and the diffuse background

\bibitem[\protect\citeauthoryear{{Tittley} \& {Henriksen}}{{Tittley} \&
  {Henriksen}}{2005}]{Tittley2005}
{Tittley} E.~R.,  {Henriksen} M.,  2005, \mn@doi [\apj] {10.1086/425952}, \href
  {http://adsabs.harvard.edu/abs/2005ApJ...618..227T} {618, 227}

\bibitem[\protect\citeauthoryear{{Vazza}, {Brunetti}, {Kritsuk}, {Wagner},
  {Gheller}  \& {Norman}}{{Vazza} et~al.}{2009}]{Vazza2009}
{Vazza} F.,  {Brunetti} G.,  {Kritsuk} A.,  {Wagner} R.,  {Gheller} C.,
  {Norman} M.,  2009, \mn@doi [\aap] {10.1051/0004-6361/200912535}, \href
  {https://ui.adsabs.harvard.edu/abs/2009A&A...504...33V} {504, 33}

\bibitem[\protect\citeauthoryear{{Venturi} et~al.,}{{Venturi}
  et~al.}{2022}]{Venturi2022}
{Venturi} T.,  et~al., 2022, \mn@doi [\aap] {10.1051/0004-6361/202142048},
  \href {https://ui.adsabs.harvard.edu/abs/2022A&A...660A..81V} {660, A81}

\bibitem[\protect\citeauthoryear{Walker \& Lau}{Walker \&
  Lau}{2022}]{Walker2022book}
Walker S.~A.,  Lau E.~T.,  2022, Cluster Outskirts and Their Connection to the
  Cosmic Web.
Springer Nature Singapore, Singapore, pp 1--37,
  \mn@doi{10.1007/978-981-16-4544-0_120-1}

\bibitem[\protect\citeauthoryear{{Walker}, {Fabian}  \& {Sanders}}{{Walker}
  et~al.}{2013}]{Walker2013_CentaurusXMM}
{Walker} S.~A.,  {Fabian} A.~C.,   {Sanders} J.~S.,  2013, \mn@doi [\mnras]
  {10.1093/mnras/stt1515}, \href
  {http://adsabs.harvard.edu/abs/2013MNRAS.435.3221W} {435, 3221}

\bibitem[\protect\citeauthoryear{{Walker}, {Fabian}  \& {Sanders}}{{Walker}
  et~al.}{2014}]{Walker2014}
{Walker} S.~A.,  {Fabian} A.~C.,   {Sanders} J.~S.,  2014, \mn@doi [\mnras]
  {10.1093/mnrasl/slu040}, \href
  {http://adsabs.harvard.edu/abs/2014MNRAS.441L..31W} {441, L31}

\bibitem[\protect\citeauthoryear{{Walker}, {Sanders}  \& {Fabian}}{{Walker}
  et~al.}{2016}]{Walker2016}
{Walker} S.~A.,  {Sanders} J.~S.,   {Fabian} A.~C.,  2016, \mn@doi [\mnras]
  {10.1093/mnras/stw1367}, \href
  {http://adsabs.harvard.edu/abs/2016MNRAS.461..684W} {461, 684}

\bibitem[\protect\citeauthoryear{{Walker}, {ZuHone}, {Fabian}  \&
  {Sanders}}{{Walker} et~al.}{2018}]{Walker2018NatAs}
{Walker} S.~A.,  {ZuHone} J.,  {Fabian} A.,   {Sanders} J.,  2018, \mn@doi
  [Nature Astronomy] {10.1038/s41550-018-0401-8}, \href
  {https://ui.adsabs.harvard.edu/abs/2018NatAs...2..292W} {2, 292}

\bibitem[\protect\citeauthoryear{{Walker}, {Mirakhor}, {ZuHone}, {Sanders},
  {Fabian}  \& {Diwanji}}{{Walker} et~al.}{2022}]{Walker2022}
{Walker} S.~A.,  {Mirakhor} M.~S.,  {ZuHone} J.,  {Sanders} J.~S.,  {Fabian}
  A.~C.,   {Diwanji} P.,  2022, \mn@doi [\apj] {10.3847/1538-4357/ac5894},
  \href {https://ui.adsabs.harvard.edu/abs/2022ApJ...929...37W} {929, 37}

\bibitem[\protect\citeauthoryear{ZuHone \& Su}{ZuHone \&
  Su}{2022}]{ZuHone2022book}
ZuHone J.,  Su Y.,  2022, The Merger Dynamics of the X-Ray-Emitting Plasma in
  Clusters of Galaxies.
Springer Nature Singapore, Singapore, pp 1--44,
  \mn@doi{10.1007/978-981-16-4544-0_124-1}

\bibitem[\protect\citeauthoryear{{ZuHone}, {Markevitch}, {Brunetti}  \&
  {Giacintucci}}{{ZuHone} et~al.}{2013}]{ZuHone2013}
{ZuHone} J.~A.,  {Markevitch} M.,  {Brunetti} G.,   {Giacintucci} S.,  2013,
  \mn@doi [\apj] {10.1088/0004-637X/762/2/78}, \href
  {http://adsabs.harvard.edu/abs/2013ApJ...762...78Z} {762, 78}

\bibitem[\protect\citeauthoryear{{Zuhone} \& {Roediger}}{{Zuhone} \&
  {Roediger}}{2016}]{Zuhone2016review}
{Zuhone} J.~A.,  {Roediger} E.,  2016, \mn@doi [Journal of Plasma Physics]
  {10.1017/S0022377816000544}, \href
  {http://adsabs.harvard.edu/abs/2016JPlPh..82c5301Z} {82, 535820301}

\makeatother
\end{thebibliography}








\bsp	
\label{lastpage}
\end{document}